# Complex Spin Hamiltonian Represented by Artificial Neural Network

Hongyu Yu,[a,b,c]* Changsong Xu,[d,a]* Feng Lou,[a] L. Bellaiche,[d] Zhenpeng Hu,[c] Xingao Gong,[a,b] Hongjun Xiang[a,b,†]

[a]*Key Laboratory of Computational Physical Sciences (Ministry of Education), Institute of Computational Physical Sciences, and Department of Physics, Fudan University, Shanghai 200433, China*
[b]*Shanghai Qizhi Institution, Shanghai 200030, China*
[c]*School of Physics, Nankai University, Tianjin 300071, China*
[d]*Physics Department and Institute for Nanoscience and Engineering, University of Arkansas, Fayetteville, Arkansas 72701, USA*

**Abstract**: The effective spin Hamiltonian method is widely adopted to simulate and understand the behavior of magnetism. However, the magnetic interactions of some systems, such as itinerant magnets, are too complex to be described by any explicit function, which prevents an accurate description of magnetism in such systems. Here, we put forward a machine learning (ML) approach, applying an artificial neural network (ANN) and a local spin descriptor to develop effective spin potentials for any form of interaction. The constructed Hamiltonians include an explicit Heisenberg part and an implicit non-linear ANN part. Such a method successfully reproduces artificially constructed models and also sufficiently describe the itinerant magnetism of bulk $Fe_3GeTe_2$. Our work paves a new way for investigating complex magnetic phenomena (e.g., skyrmions) of magnetic materials.

**Main text**

Magnetism covers a major area of condensed matter physics and is of both fundamental and technical importance. Novel magnetic phases, such as skyrmion states, spin liquids and spin glasses, have been attracting a lot of scientific attention in recent years [1–8]. These diverse spin textures and colorful physics are driven by magnetic interactions of different forms. Effective spin Hamiltonian containing necessary interactions is a low-energy approximation that is well-adopted to describe the magnetic properties of various systems [9–11]. A well-known example is the Heisenberg model, $H_{\text{spin}} = \sum_{i>j} J_{ij} \mathbf{S}_i \cdot \mathbf{S}_j$, where $J_{ij}$ denotes the coupling strength between spins $\mathbf{S}_i$ and $\mathbf{S}_j$, and it can be obtained from *ab initio* calculations or fitting to experiments. Such a model is frequently used since it captures the basic physics of ferromagnetism and antiferromagnetism, as well as, frustrations in most systems. However, such a model becomes insufficient when novel mechanisms come into play. For example, novel states arise from Dzyaloshinskii–Moriya interaction (DMI) [12] and Kitaev interaction [13] that is related to spin-orbit coupling (SOC), and it is also challenging to describe systems in which many-body and higher-order interactions are crucial even if they do not involve SOC. For instance, (i) biquadratic interactions are reported to widely exist in 2D magnets [14,15], (ii) three/four-body interactions are found to stabilize some skyrmions [16], and (iii) topological orbital magnetism with even higher orders are predicted to play important roles in non-coplanar spin patterns [17]. There may exist numerous different many-body and higher-order interactions, which makes it difficult to numerically determine the important ones and their strengths, especially when many of such interactions work collaboratively. Furthermore, the forms of couplings among itinerant moments of metals can be elusive and unprecedentedly complicated, which cannot be described by common Hamiltonians [18]. Such facts indicate that a new method for constructing an effective spin Hamiltonian for *any* system is highly desired to accurately describe complex magnetism.

The fast-developing machine learning potential (MLP) method is a promising solution. MLP built with artificial neural networks (ANN) has demonstrated its significant advantages in accuracy and efficiency [19–25]. It has been proven to be a powerful tool to solve problems in simulations of materials [19,26–28]. Descriptors are the suitable input representation of atomistic systems for machine learning [29]. To the best of our knowledge, all descriptors, as the input of ANN, generally do not consider the intrinsic property, spin, which dominates the behavior of the magnetic systems. Descriptors without information of spin fail to describe the spin interactions in magnetic materials. Therefore, spin descriptors and models used in magnetic materials are necessary to describe the magnetic information with the spin distribution. The application of MLPs to spin systems has just started and is still in its infancy. Recent attempts include (i) the magnetic moment tensor potentials (mMTP) that can well describe the single element magnetic systems [32] and (ii) the spin-dependent atom-centered symmetry functions (sACSF) [28]. However, both methods only deal with collinear magnetic orders, which prevent the study of complex noncollinear states (e.g., spirals, skyrmions, and bimerons). Hence, a general MLP method dedicated to spin systems with complex interactions and colorful spin patterns is still highly desired.

In this Letter, we present a general machine learning method to construct spin Hamiltonians based on ANN. Such a method adopts local spin descriptors, where spins can freely rotate to any three-dimensional direction. As we will show, the obtained Hamiltonian contains an explicit Heisenberg part and an implicit nonlinear part, which gather all possible non-SOC interactions forms. Moreover, such a method can not only reproduce any artificially created models with peculiar complex interactions but also can well describe the complex magnetism of $Fe_3GeTe_2$, where ferromagnetism, diverse spirals and skyrmionic states are observed [30–32].

*Machine learning spin Hamiltonian approach*. In this work, we aim at building the machine learning spin Hamiltonian (MLSH) $H$ for a given atomic structure, which can be subsequently used to obtain the thermodynamic and kinetic properties of the spin system by performing Monte-Carlo simulations or spin dynamic simulations. In principle, one can express the MLSH using solely ANN. However, we find that it is beneficial (see tests below) if the MLSH contains explicitly the isotropic Heisenberg spin Hamiltonian besides the ANN part: $H = H_{HB} + H_{NN}$, where the Heisenberg part $H_{HB} = \sum_{i>j} J_{ij} \mathbf{S}_i \cdot \mathbf{S}_j$ ($J_{ij}$ are the exchange interactions and $\mathbf{S}_i$ are spin vectors) and $H_{NN}$ is the remaining energy contribution as described by the ANN. The Heisenberg interaction is of the second-order two-body form and usually dominates in most magnets. However, many-body and higher-order spin interactions are important in many interesting systems such as itinerant metallic magnets. The ANN contribution $H_{NN}$ is intended to describe such complicated interactions. Such treatment of separating Hamiltonian into explicit Heisenberg and implicit ANN parts results in a good balance between efficiency and accuracy. The details (including the training of ANN) of building the MLSH are presented in the SM. Below we will describe some key issues when constructing spin Hamiltonians that require special treatments.

Similar to the case of machine learning atomic potentials [19,33,34], the ANN part of the spin Hamiltonian can be written as a sum of contributions from individual central spins:

$$H_{NN} = \sum_i H_{NN,i}(\boldsymbol{D}_i), \qquad \text{Eq. (1)}$$

where $\boldsymbol{D}_i$ is the descriptor vector that describes the local spin environment of the *i*-th central spin, and $H_{NN,i}$ is a function expressed in terms of ANN. A natural way to choose central spins in $H_{NN}$ is that each spin in the system acts as a central spin. The form (*i.e.*, ANN parameters) of $H_{NN,i}$ depends on the local atomic environment around the *i*-th central spin. As inequivalent Wyckoff positions have different local atomic environments, we can classify the central spin according to the Wyckoff position that it belongs to. If the crystal has *n* inequivalent magnetic Wyckoff positions, the ANN part of the Hamiltonian can be written as $H_{NN} = \sum_{\mu=1}^{n} \sum_{\gamma} H_{NN,\mu}(\boldsymbol{D}_{i=[\mu,\gamma]})$, where $\mu$ and $\gamma$ are the spin class index and index of spins within the $\mu$-th spin class, respectively.

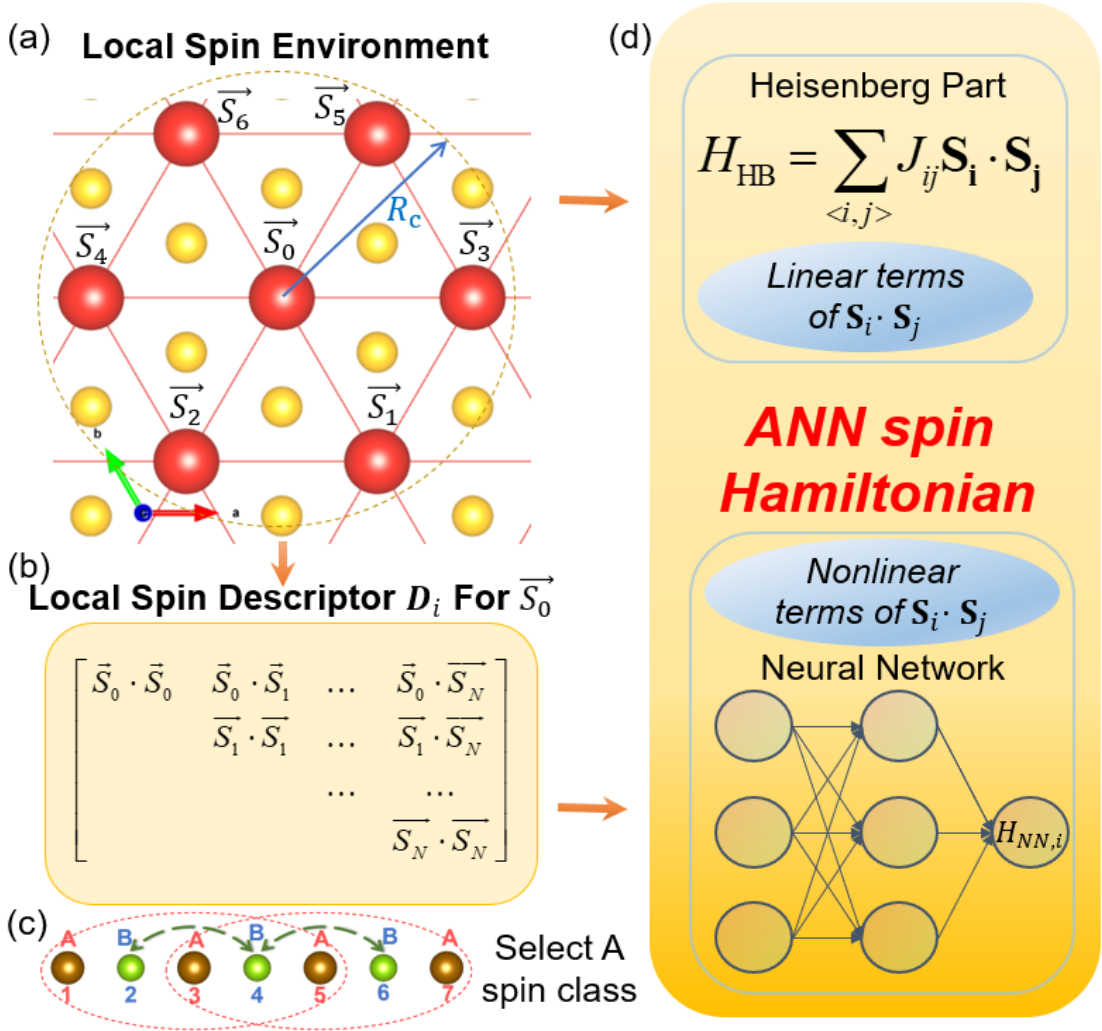

**Figure 1.** Workflow of the present ANN method to obtain spin Hamiltonians. (a) Illustration of the local structure centered on spin $\vec{S}_0$ within a cutoff radius of $R_c$. The Cu$M$O$_2$ structure is shown here as an example, where the red balls indicate magnetic sites of $M$. (b) The local spin descriptor centered on spin $\vec{S}_0$, involving all neighboring spins within cutoff radius $R_c$. (c) Demonstration that one set of NN is enough for two different spin classes, as the descriptor of A spin class includes the spin interaction information of the B site. (d) The whole Hamiltonian that contains an explicit Heisenberg part and an implicit part, which includes all other forms of interactions. The total energy consists of energies from both parts.

Let us now concentrate on the spin descriptors $D_i$ that depends not only on central spin $S_i$ itself, but also on the $N_i$ neighboring spin vectors of $S_i$. The reason why we need to transform the spin vectors into descriptors instead of directly using the spin vectors as inputs to the ANN is that the symmetry properties of the spin Hamiltonian can be enforced strictly through the descriptors. In this work, we mainly focus on the usual case where the spin-orbit coupling (SOC) effects can be neglected. To consider the SOC effects within our framework (*e.g.*, DMI, Kitaev interaction and/or single ion anisotropy terms), we can add these terms into the explicit Heisenberg part. The ANN part of the Hamiltonian $H_{NN}$ is intended to deal with the high-order spin interactions in the absence of SOC effects. In the absence of SOC effects, the spin interaction energy is invariant under any simultaneous global rotation of all the spin vectors. In addition, the spin interaction energy is invariant under the time-reversal operation that reverses all the spin directions. To satisfy these conditions, each element of the descriptor vector is set to a dot product between two spin vectors $\mathbf{S}_p \cdot \mathbf{S}_q$. To be more specific, let us consider the spin descriptor $\mathbf{D}_0$ of the central spin 0 with $N_0$ spin neighbors within the cutoff distance $R_c$ (see Fig. 1). We define the descriptor vector $\mathbf{D}_0$ with the elements taken from the upper triangular matrix of $(\mathbf{S}_p \cdot \mathbf{S}_q)_{N_0 \times N_0}$: $\mathbf{D}_0 = (\mathbf{S}_0 \cdot \mathbf{S}_0, ..., \mathbf{S}_0 \cdot \mathbf{S}_{N_0}, \mathbf{S}_1 \cdot \mathbf{S}_1, ..., \mathbf{S}_1 \cdot \mathbf{S}_{N_0}, ..., \mathbf{S}_{N_0} \cdot \mathbf{S}_{N_0})^T$. There is another symmetry that the descriptor must be able to account for. If the Wyckoff position that the central spin belongs to displays the point group $G$ symmetry with $m$ operations, then any operation $R$ of $G$ on the spin configuration $(\mathbf{S}_0, \mathbf{S}_1, ..., \mathbf{S}_{N_0})$ will leave the spin interaction energy invariant. Note that $R$ will not only change the spin directions but may also result in a permutation of the spin sites. To take the point group symmetry into account, we first generate all $m$ spin configurations by applying $m$ operations to a given spin configuration, then we select a representative spin configuration among the $m$ spin configurations according to some predefined criteria (see SM). The representative spin configuration is used to construct the spin descriptor $\mathbf{D}_0$. In this way, these $m$ spin configurations will be guaranteed to have the same spin interaction energy. To better demonstrate the process of constructing the spin descriptor, a simple example is presented in SM.

In our above discussion on the ANN part of the spin Hamiltonians (see Eq. 1), every spin in the system acts as a central spin. We find (see tests below) that the efficiency can be significantly improved by reducing the number of central spins in the case of multiple inequivalent magnetic Wyckoff positions (i.e., $n > 1$). Consider the crystal [see Fig. 1(c)] with two inequivalent magnetic Wyckoff positions A and B, one can select A as the central spin but leave out B. This is because the spin interactions involving the B spin are included in that related to the neighboring central spins (A spins here) within a cutoff distance $R_c$. Reducing the number of central spins not only reduces the number of ANN parameters and thus the risk of overfitting, but also speeds up the evaluation of the total spin interaction energies. We note that this idea of using a reduced number of central sites can also be adopted to construct the highly efficient machine learning atomic potential for simulating atomic process (*e.g.*, conduction of heat, ferroelectric phase transition) that maintains the basic framework of the crystal (see SM).

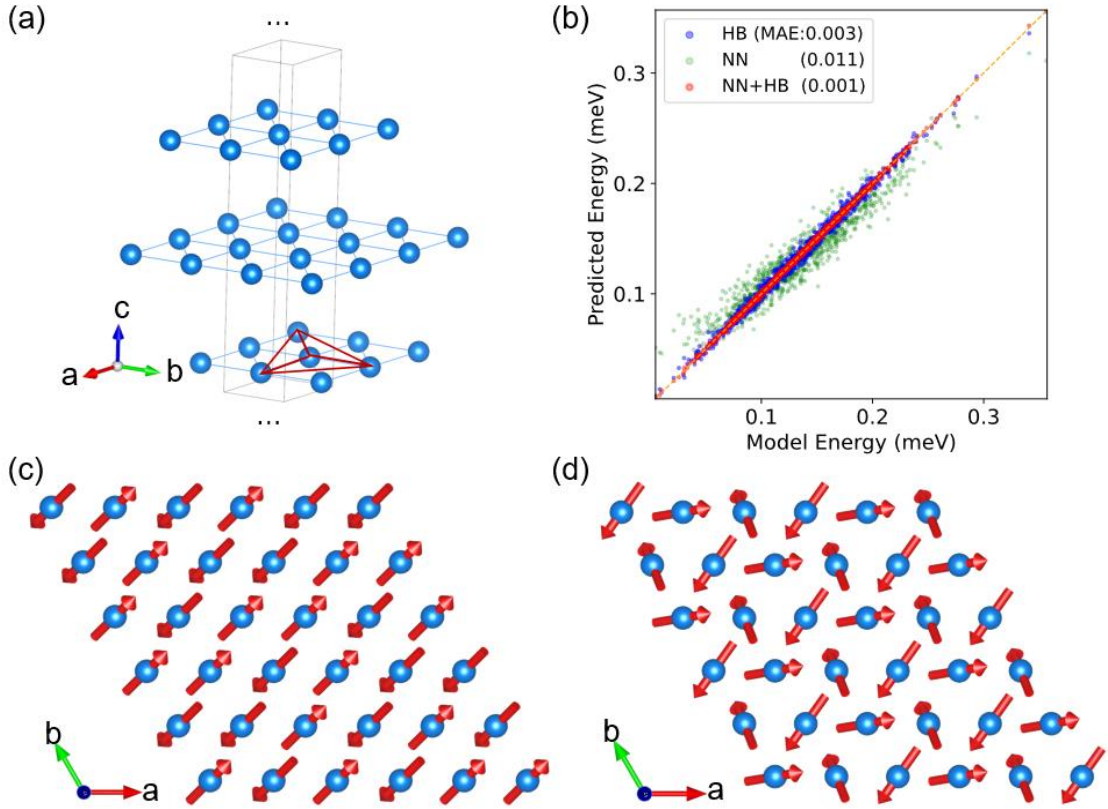

**Figure 2.** (a) Schematization of the Cu$M$O$_2$-type structure, where only the magnetic $M$ sites are shown. The atoms connected by red lines form an example of the four-body cluster, for which the fourth-order interaction is considered in Eq. (2). (b) Comparison of the energies among three different fitting methods – only Heisenberg terms, only NN and both Heisenberg and NN. The Mean Absolute Error (MAE) of each fitting method is shown in the legend. (c) The (↑↑↓↓) AFM magnetic ground state of the exact model Hamiltonian is reproduced by the HB+NN method. (d) The frustration induced a short

period spiral, which is the ground state predicted by the pure Heisenberg method.

*An artificial model*. To validate our machine learning approach for constructing spin Hamiltonian, we employ a simple model Hamiltonian. This model is intended to describe the spin interactions between magnetic ions that occupy the Al sites of the $CuAlO_2$-type [35] structure. The model spin Hamiltonian contains both intralayer interactions within the triangular lattice and interlayer couplings:

$$H = \sum_{<ij>_n}^{n=1,\ldots,5} J_n \mathbf{S}_i \cdot \mathbf{S}_j + \sum_{<ij>_n}^{n=1,5,7} K_n \left(\mathbf{S}_i \cdot \mathbf{S}_j\right)^2 + \sum_{<ijkl>_1} L_1 \left(\mathbf{S}_i \cdot \mathbf{S}_j\right)(\mathbf{S}_k \cdot \mathbf{S}_l). \qquad (2)$$

Here (i) $J_n$ are the 2nd-order Heisenberg spin interactions, where $J_1 - J_3$ are AFM intralayer couplings and are thus strongly frustrated, while $J_4$ are relatively weak interlayer couplings and $J_5$ is set to zero but included while fitting for testing purposes; (ii) $K_n$ are the 4th-order biquadratic interactions for the 1st, 5th and 7th nearest neighboring pairs; and (iii) $L_1$ represents the four-body fourth-order interaction for the triangular cluster with the fourth site in the center of the triangle as shown in Fig. 2(a). The values of $J, K$, and $L$ as listed in Table I, are set in such a way that leads to a complex spin model with strong frustrations and high-order interactions. Our MC simulations using the exact model Hamiltonian suggest that our model with the parameters from Table I results in a nontrivial (↑↑↓↓) antiferromagnetic (AFM) ground state, as shown in Fig. 2(c).

We now try to reproduce the results from the model Hamiltonian by building our MLSH. The data set of 10,000 fully random spin configurations is generated based on a 4×4×2 supercell with 96 magnetic sites (*i.e.*, six layers of 4×4 triangular lattices), and their energies are obtained from Eq. (2). Our MLSH contains five Heisenberg terms and a multi-layer fully connected neural network. During the fitting, 90% of the data are used for training and the remaining 10% for testing. To check whether the explicit inclusion of the Heisenberg interaction is crucial to the accuracy of the MLSH, we consider three different forms of the spin Hamiltonian: (i) HB Hamiltonian which only includes the Heisenberg interactions (five terms for this system) (ii) NN Hamiltonian which includes the ANN part, and (iii) NN+HB full Hamiltonian which includes both Heisenberg and ANN parts.

As shown in Fig. 2(b), the comprehensive NN+HB Hamiltonian is found to display a much smaller ($1.0 \times 10^{-5}$ meV/site) mean absolute error (MAE, defined as $\frac{\sum |f(x_i) - y_i|}{n}$) in total energy than that ($3.4 \times 10^{-5}$ meV/site) of the HB Hamiltonian and that ($1.1 \times 10^{-4}$ meV/site) of the NN Hamiltonian. Further, by performing Monte Carlo (MC) simulations, we find that the NN+HB Hamiltonian can correctly reproduce the (↑↑↓↓) AFM ground state [see Fig. 2(c)], while the HB Hamiltonian predicts incorrectly a non-collinear ground state [see Fig. 2(d)]. The Néel temperature obtained with the NN-HB Hamiltonian is also in agreement with that obtained with the exact model Hamiltonian. Note that the MLSH can also well describe non-collinear exotic spin states. For example, by tuning $J_1$ from 3.659 meV to 18.295 meV, the ground state of the exact model becomes a noncollinear order spiral state with q-vector as (1/3, 0) in triangular lattice plane, which can be well reproduced by the MLSH approach.

**Table I**. Original and fitted magnetic coefficients for the artificial model in Eq. (1) and the real compound of $Fe_3GeTe_2$ (Energy unit meV). The subscript $i, j$ of bulk $Fe_3GeTe_2$ part represents the atom index of the two-body Heisenberg interaction $J_{i,j}$ in Fig. 1a. Note that "HB" denotes the results from pure Heisenberg model, while "NN+HB" refers to the model with Heisenberg part and implicit part treated by ANN separately.

|  | Artificial model Eq. (1) | | | | Bulk $Fe_3GeTe_2$ | |
| --- | --- | --- | --- | --- | --- | --- |
|  | Model | NN+HB | HB |  | NN+HB | HB |
| $J_1$ | 3.659 | 3.714 | 3.659 | $J_{1,3}$ | -0.486 | -0.474 |
| $J_2$ | 1.117 | 0.988 | 1.114 | $J_{1,2}$ | -0.207 | -0.199 |
| $J_3$ | 1.728 | 1.779 | 1.729 | $J_{2,2}$ | 0.069 | 0.065 |
| $J_4$ | 1.003 | 0.995 | 0.998 | $J_{1,1}$ | 0.034 | 0.023 |
| $J_5$ | 0.000 | -0.002 | 0.003 | $J_{1,3(2)}$ | -0.078 | -0.069 |
| $K_1$ | -1.386 | \ | \ | $J_{1,6}$ | 0.007 | 0.010 |
| $K_5$ | -0.376 | \ | \ | $J_{2,6}$ | -0.008 | -0.006 |
| $K_7$ | -0.143 | \ | \ | $J_{2,5}$ | -0.006 | -0.007 |
| $L_1$ | 0.316 | \ | \ |  |  |  |

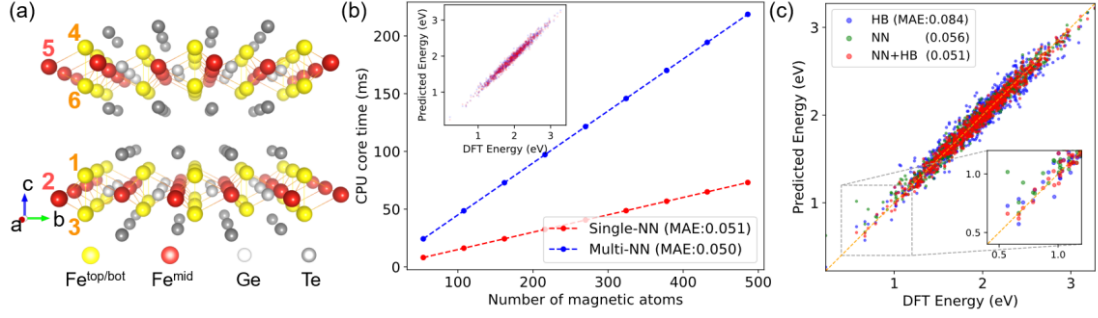

**Figure 3.** (a) Crystal structure of bulk $Fe_3GeTe_2$. The unit cell consists of two Van der Waals layers with six Fe atoms, that are categorized into two types, two $Fe^{mid}$ and four $Fe^{top}/Fe^{bot}$. (b) Comparison of the CPU time between the single-NN method and two-NN method, as a function of magnetic atom number. Both sets of simulations are performed on Intel Xeon Gold 6244. The inset indicates similar accuracy of the two methods. (c) Comparison among energies from DFT and three different fitting strategies. The MAE of each method is shown in the legend.

*Metallic ferromagnet $Fe_3GeTe_2$*. $Fe_3GeTe_2$ (FGT) is a metallic Van der Waals layered ferromagnetic (FM) material [see crystal structure in Fig. 3(a)] with a Curie temperature of about 300 K [36]. Recently, FGT thin films of about a few tens of nanometers in thickness are intensively studied and skyrmions and domain walls are widely observed there [31,32,37–39]. These interesting observations suggest that complex high-order interactions are essential besides second-order Heisenberg interactions [36,40]. We thus apply the presently developed ANN method to explore the magnetism of bulk FGT.

As shown in Fig. 3(a), each Van de Waals layer of bulk FGT contains three layers of Fe ions, where the top and bottom Fe ions are equivalent but are different from the middle Fe ion. In another word, there are two kinds of inequivalent magnetic Wyckoff positions, *i.e.*, $n = 2$. To see whether the strategy of reducing the number of central spins works, we consider two cases: (i) use all Fe sites as central spins (refers to as multi-NN strategy) *versus* (ii) only use the middle Fe sites as central spins (refers to as single-NN strategy). For our NN-HB Hamiltonian, the Heisenberg part includes both intralayer and interlayer couplings interactions up to 18 Å, while the ANN part only considers intralayer interactions with $R_c$ = 9 Å. The data set consists of 8000 random spin configurations within a 3×3×1 supercell, which contains 54 Fe, and their energies are computed with DFT (see Methods in Supplementary Materials). Again, 90% of the data are used for training and 10% for testing. We find that the MLSH obtained with the single-NN strategy is as accurate as that with the multi-NN strategy [see insets of Fig. 3(c)]. However, the single-NN strategy is much more efficient than the multi-NN strategy for two reasons. In the training process, it takes much less time as the number of parameters in the single-NN strategy is one half of that in the multi-NN strategy. When using the MLSH, the calculations of the spin interaction energies in the case of single-NN strategy costs only 1/3 computational time of that in the case of the multi-NN strategy because the number of central spins in the formal case is 1/3 of that in the latter case [see Fig. 3(c)]. Hereafter, we will thus adopt the more efficient single-NN strategy for FGT.

Similar to the case of the model spin Hamiltonian, we find that the NN+HB Hamiltonian is more accurate than the HB Hamiltonian and NN Hamiltonian: the MAE in these three cases are 0.94 meV/Fe, 1.56 meV/Fe, and 1.04 meV/Fe, respectively; in addition, the errors in the predicted energies for some low-energy spin configurations are rather large when using the NN Hamiltonian [see insets of Fig. 3(b)]. These tests suggest that the NN+HB Hamiltonian is desirable in general cases.

The finite-temperature properties of FGT are then simulated with parallel tempering Monte Carlo (PTMC) [41,42] (see PTMC details in SM). The ANN model from NN+HB scheme (iii) results in a Curie temperature of 554 K. Such value is based on the data from DFT with no Hubbard $U$ correction, while further calculations (See details in SM) indicate that finite $U$ correction may lead to Curie temperature that compares well with experiment [43].

In summary, we develop a machine learning (ML) method to construct spin Hamiltonians. Such a method involves a local spin descriptor and adopts an artificial neural network to describe the complex spin interactions of any form. During the fitting, the Heisenberg part and other interactions are treated separately, which leads to improvement in accuracy. The effectiveness and accuracy of such a method are verified with an artificial complex Hamiltonian and also the case of itinerant magnet $Fe_3GeTe_2$. The presently developed method is thus suitable to describe complex magnetic properties of materials that are hard to be captured by selected simple forms of interactions.

*Acknowledgements*. The work at Fudan is supported by NSFC (11825403, 11991061). CX and LB thank the Vannevar Bush Faculty Fellowship Grant No. N00014-20-1-2834 from the Department of Defense.

\* Contributed equally to this work.

† hxiang@fudan.edu.cn